\documentclass[prl,aps,twocolumn,floatfix]{revtex4}
\usepackage{epsfig} 
\usepackage{graphicx} 

\begin{document} 
\title{Scrambled Mean Field Approach to the Quantum Dynamics of Degenerate Bose Gases}

\author{Igor E. Mazets} 
\affiliation{
Vienna Center for Quantum Science and Technology, Atominstitut, TU~Wien,~Stadionallee~2,~1020~Vienna,~Austria; \\
Wolfgang Pauli Institute c/o Fakult\"{a}t f\"{u}r Mathematik,
Universit\"{a}t Wien, Oskar-Morgenstern-Platz 1, 1090 Vienna, Austria}

\begin{abstract} 
We present a novel approach to modeling dynamics of trapped, degenerate, 
weakly interacting  Bose gases beyond the mean field limit. 
We transform a many-body problem to the interaction representation with respect to 
a suitably chosen part of the Hamiltonian and only then apply a multimode coherent-state \textit{ansatz}. 
The obtained equations are almost as simple as the Gross--Pitaevskii equation, but our approach captures essential 
features of the quantum dynamics such as the collapse of coherence.  
\end{abstract}
\maketitle

The mean-field approximation has become a powerful tool for modeling dynamics of degenerate gases of weakly interacting bosonic atoms 
\cite{Dalfovo,ufn98,Leggett-rev}. In this approximation, a Bose--Einstein condensate (BEC) or, in a case of low-dimensional 
geometry, a quasicondensate is described by a complex-valued classical field 
subject to the time-dependent Gross-Pitaevskii equation (GPE). Thermal and even quantum zero-point fluctuations can be 
incorporated into this classical-field picture within the truncated Wigner approximation (TWA) \cite{TWA1} via 
initial conditions. Unfortunately, the TWA with quantum noise 
provides physically meaningful solutions on rather a restricted time scale \cite{TWA2}. 
An alternative to the mean-field calculations is given by the quantum Boltzmann equation that can be derived by the standard 
techniques of the non-equilibrium quantum field theory \cite{Gasenzer1,Werner,Gasenzer2}. Buchhold and Diehl \cite{DiehlEPJD} 
derived kinetic equations not only for populations, but also for anomalous correlations in phonon modes in one dimension (1D). 
However, the quantum field theory methods are developed for a bulk medium in the thermodynamic limit, where the 
excitation spectrum is continuous and on-shell self-energies have therefore a non-zero imaginary part. In experiment, 
the finite size of trapped ultracold atomic clouds makes their excitation spectra discrete with frequencies of different excitation modes 
being well resolvable \cite{Jin96,Mewes96}. The multiconfigurational time-dependent Hartree method for bosons (MCTDHB) \cite{Alon08} 
is suitable for numerical modeling the non-mean-field dynamics of finite-size systems. However, it seems that the MCTDHB 
well describes the experimental data only in situations where the number of involved configurations remains small because of 
limitations specific for a given process, such as parametric excitation of a Bose--Einstein condensate (BEC) \cite{Hulet19}, 
and  remains  otherwise  suitable mainly for few-body problems. A recently developed 
truncated conformal space approach \cite{Takacs2,Takacs1} works well at relatively low excitation energies of a bosonic system, 
as numerical diagonalization methods  in general do. 

Experiments with ultracold bosonic gases exhibiting effects beyond the mean field include, first of all, observations of the collapse and 
revival dynamics in optical lattices \cite{Bloch1,Bloch2}. Moreover, redistribution of atomic population between lattice sites accompanying 
this phenomenon has been detected in a recent experiment  \cite{Zhou}. The available theory \cite{phase-diff,Imamoglu-theor,Kuklov} 
does not account for multimode aspects of the problem and remains a matter-wave analog of the well-known Jaynes--Cummings model 
in quantum optics \cite{cum-collapse,eber-collapse}. 

The multimode approach that we present here can be called a scrambled mean field method. It 
bears certain similarities to the rotated Hartree method \cite{Cederbaum87},  
but is remarkably simpler. 
Consider a Hamiltonian 
$\hat H=\hat H_\mathrm{h}+\hat H_\mathrm{a}$ that describes a BEC or a 
low-dimensional quasicondensate in collective variables \cite{Dalfovo,ufn98,Leggett-rev}. 
Its harmonic part, $\hat H_\mathrm{h}$, can be written after diagonalization as 
$\hat H_\mathrm{h}=\hbar \sum _j\omega _j\hat b_j^\dag \hat b_j$, 
where $\hat b_j^\dag $ and $\hat b_j$ are the creation and annihilation operators, 
respectively, of excitation quanta in the $j$th elementary mode with the fundamental frequency $\omega _j$ ($\hat b_j^\dag $ and $\hat b_j$ 
obey the standard bosonic commutation rules). 
We denote the eigenstates of $\hat H_\mathrm{h}$, i.e., Fock states of 
elementary excitations, by  $|\mathbf{n}\rangle =|\{ n_1,\, n_2,\, n_3,\, \dots \} \rangle $, where 
$\hat b_j^\dag \hat b_j |\mathbf{n}\rangle =n_j|\mathbf{n}\rangle $. 
The anharmonic part $\hat H_\mathrm{a}$ of the Hamiltonian  describes interaction between elementary excitations. 
This interaction is assumed to be small in order to make elementary excitations well defined. 
Usually $\hat H_\mathrm{a}$ can be expanded in Taylor series in $\hat b_j^\dag $ and $\hat b_j$ 
beginning from a cubic term in a general case. 
The first-order perturbative correction to the energy of the state $|\mathbf{n}\rangle $ is given by the matrix element 
$\langle \mathbf{n}|\hat H_\mathrm{a}|\mathbf{n}\rangle $. The lowest order term that contributes to 
$\langle \mathbf{n}|\hat H_\mathrm{a}|\mathbf{n}\rangle $ 
is a quartic one, more precisely, its diagonal part  $\hat H_\mathrm{qd} =\frac \hbar 2 \sum _{j,j^\prime } g_{jj^\prime } \hat b^\dag _j 
\hat b^\dag _{j^\prime }\hat b_{j^\prime }\hat b_j$, so that $\langle \mathbf{n}|\hat H_\mathrm{a}|\mathbf{n}\rangle = \frac \hbar 2 
\sum _{j,j^\prime }g_{jj^\prime }n_j(n_{j^\prime } -\delta _{jj^\prime } )+\Delta E_\mathbf{n}$, where $\Delta E_\mathbf{n}$ 
is the contribution of higher-order terms and $g_{jj^\prime }=g_{j^\prime j}$.

Now we need to introduce a unitary operator that induces quantum correlations between the modes. In contrast to Ref. \cite{Cederbaum87}, 
it contains no time-dependent parameters to be determined from the variational principle, but we  
derive instead its form from the perturbation theory considerations. 
We rearrange the terms in the Hamiltonian as $\hat H=\hat H_0+\hat V$, where $\hat H_0 =\hat H _\mathrm{h}+\hat H_\mathrm{qd}$ and  
$\hat V=\hat H _\mathrm{a}-\hat H_\mathrm{qd}$. The evolution of the wave function $|\Psi \rangle $ os the system is governed by the 
Schr\"odinger equation $i\hbar \frac \partial {\partial t} |\Psi \rangle =\hat H|\Psi \rangle $. Our first step is to introduce 
the interaction representation with respect to the diagonalizable anharmonic Hamiltonian $\hat H_0$. This is done by an unitary 
transformation $|\Psi \rangle = \hat U_\mathrm{r}(t)|\tilde \Psi \rangle $, where  $\hat U_\mathrm{r}(t)=\exp ( -i\hat H_0t/\hbar )$. 
This transformation does not mix different Fock states of elementary excitations, but induces correlations between modes via the 
energy shift for elementary excitations  depending on the quantum state of all the other modes and 
hence ``scrambles" $|\tilde \Psi \rangle $. After this transformation  the Schr\"odinger equation reads as 
\begin{equation} 
i\hbar \frac \partial {\partial t} |\tilde \Psi \rangle =\hat{\tilde V}(t)|\tilde \Psi \rangle , 
\label{ro.1}
\end{equation} 
where 
\begin{equation} 
\hat{\tilde V}(t)=\hat U_\mathrm{r}^{-1}(t)\hat V\hat U_\mathrm{r}(t). 
\label{ro.2} 
\end{equation} 
The field operators for the modes transform as  
\begin{equation} 
\hat U_\mathrm{r}^{-1}(t)\hat b_j\hat U_\mathrm{r}(t)=\hat{\cal A}_j \hat b_j, ~~~
\hat U_\mathrm{r}^{-1}(t)\hat b_j^\dag \hat U_\mathrm{r}(t)=\hat b_j ^\dag \hat{\cal A}_j^\dag , 
\label{ro.3} 
\end{equation}
\begin{equation} 
\hat{\cal A}_j=\exp \bigg( -i\omega _jt -i \sum _{j^\prime } g_{jj^\prime }\hat b^\dag _{j^\prime }\hat b_{j^\prime }t\bigg) . ~
\label{ro.4} 
\end{equation}   
Eqs. (\ref{ro.3}, \ref{ro.4}) allow us to write $\hat{\tilde V}$ in explicit form. 

We assume that initially, at $t=0$, the state of the system is a product of coherent states 
(normalized to unity eigenstates of the respective annihilation operators \cite{Glauber}) for each mode. This type of 
initial conditions is also assumed in the mean field theory. Now we make the variational \textit{ansatz} in the coherent state form 
not for $|\Psi \rangle $, but for the wave function in the interaction representation:  
\begin{equation} 
|\tilde \Psi \rangle = \prod _j \left[ e^{-|\psi _j|^2/2}\sum _{n_j=0}^\infty \frac {(\psi _j\hat b_j^\dag )^{n_j}}{n_j!}\right] 
|\mathbf{0}\rangle , 
\label{ro.5}
\end{equation} 
where $|\mathbf{0}\rangle $ is the vacuum state for all the modes. By minimizing the action ${\cal S}= 
\int dt\, \langle \tilde \Psi |\left(  i\hbar \frac \partial {\partial t} -\hat{\tilde V}\right) |\tilde \Psi \rangle $ we obtain 
the evolution equations for the complex functions $\psi _j(t)$ parametrizing the coherent states in Eq.~(\ref{ro.5}): 
\begin{equation} 
i\hbar \frac \partial {\partial t} \psi _j = \frac \partial {\partial \psi _j^*} \langle  \tilde \Psi |\hat{\tilde V}(t) 
|\tilde \Psi \rangle . 
\label{ro.6} 
\end{equation} 

To give a recipe for the calculation of $\langle  \tilde \Psi |\hat{\tilde V} |\tilde \Psi \rangle $, 
we assume that $\hat V$ can be expanded in Taylor series in creation and annihilation operators and consider  a term   
$\hat V^{j_1 \dots j_l}_{j^\prime _1 \dots j^\prime _{l^\prime }}=
h^{j_1 \dots j_l}_{j^\prime _1 \dots j^\prime _{l^\prime }} \prod _{\kappa =1}^l\hat b_{j_\kappa }^\dag 
\prod _{\kappa ^\prime =1}^{l^\prime }\hat b_{j^\prime _{\kappa ^\prime }}$, 
where $h^{j_1 \dots j_l}_{j^\prime _1 \dots j^\prime _{l^\prime }}$ 
is a constant.   A straightforward calculation based on elementary properties of coherent states yields 
\begin{eqnarray} 
\langle \tilde \Psi |\hat{\tilde V}^{j_1 \dots j_l}_{j^\prime _1 \dots j^\prime _{l^\prime }}|\tilde \Psi \rangle &\equiv & 
\langle \tilde \Psi |\hat U_\mathrm{r}^{-1}(t)\hat V^{j_1 \dots j_l}_{j^\prime _1 \dots j^\prime _{l^\prime }}
\hat U_\mathrm{r}(t)|\tilde \Psi \rangle 
\nonumber \\ 
&=&h^{j_1 \dots j_l}_{j^\prime _1 \dots j^\prime _{l^\prime }}\prod _{\kappa =1}^l \psi _{j_\kappa }^* 
\prod _{\kappa ^\prime =1}^{l^\prime }\psi _{j^\prime _{\kappa ^\prime }}
e^{i\nu ^{j_1 \dots j_l}_{j^\prime _1 \dots j^\prime _{l^\prime }} t} \times  
\nonumber \\  && 
\exp \bigg[ \! -\sum _j |\psi _j|^2 \Big( 1- e^{iG^{j_1 \dots j_l}_{j^\prime _1 \dots j^\prime _{l^\prime };j}t } \Big) \bigg] 
,~~~~
\label{ro.7}
\end{eqnarray} 
where 
\begin{eqnarray} 
\nu ^{j_1 \dots j_l}_{j^\prime _1 \dots j^\prime _{l^\prime }}&=&
\sum _{\kappa =1}^l\bigg( \omega _{j_\kappa }+\sum _{\lambda =\kappa +1}^l g_{\kappa \lambda }\bigg) - \nonumber \\ && 
\sum _{\kappa ^\prime =1}^{l^\prime }\bigg( \omega _{j_{\kappa ^\prime }^\prime }+
\sum _{\lambda ^\prime =\kappa ^\prime +1}^{l^\prime } g_{\kappa ^\prime \lambda ^\prime }\bigg)     , 
\label{ro.8} \\
G ^{j_1 \dots j_l}_{j^\prime _1 \dots j^\prime _{l^\prime };j}&=&\sum _{\kappa =1}^l g_{jj_\kappa } - 
\sum _{\kappa ^\prime =1}^{l^\prime } g_{jj_{\kappa ^\prime }^\prime } .  
\label{ro.9} 
\end{eqnarray}   

The number of the system modes to be taken into account is to be determined in each particular case from physics considerations. 
A good guidance can be obtained from the thermalization argument. We calculate the mean energy $E=\langle \Psi |\hat H|\Psi \rangle $ 
at $t=0$, assume that the system equilibrates at $t\rightarrow \infty $, and calculate the temperature that corresponds to the 
internal energy $E$ (if the total number of elementary excitation is conserved, we need to determine also the chemical potential). 
The number of modes with the mean number of quanta larger than 1 at thermal equilibrium will give an estimation 
for the minimal number of modes to be considered.  

After solving Eq. (\ref{ro.6}) with the initial conditions $\psi _j(0) =\psi _{0j}$, we can the find quantum-mechanical expectation value 
$\langle \hat {\cal O}\rangle \equiv \langle \Psi |\hat {\cal O}|\Psi \rangle $ for any observable $\hat {\cal O}$ as a function of time. 
For example, $\langle \hat b_j+\hat b_j^\dag \rangle =\psi _j \exp \{ -i\omega _jt-\sum _{j^\prime }|\psi _{j^\prime }|^2 [1-
\exp (i g_{jj^\prime }t)] \} +\mathrm{c.c.}\, $. 

\begin{figure}[t] 
\begin{center} 
\includegraphics[width=\columnwidth]{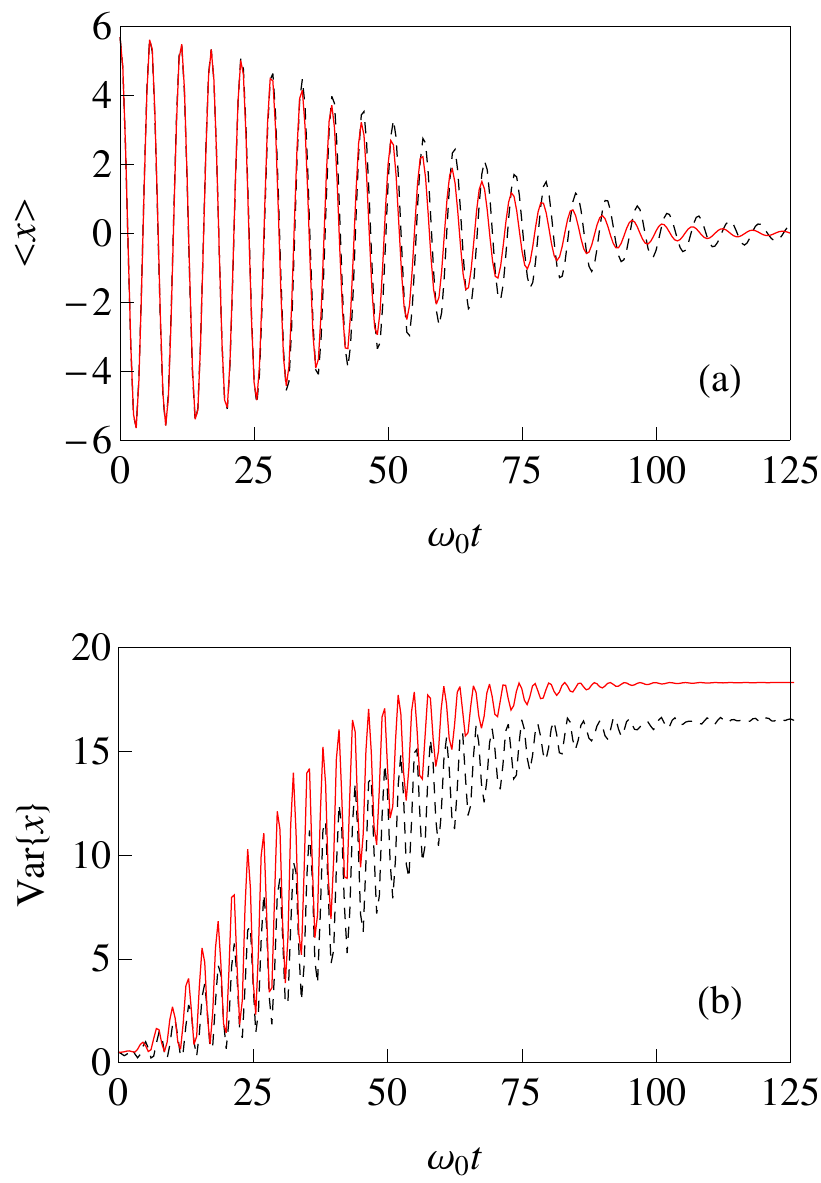} 
\end{center}

\begin{caption} 
{(Color online) (a) The mean value and (b) the variance of the co-ordinate $x$ obtained from the numerical solution  
of Eq.~(\ref{ro.6}) (red solid line) and from exactly propagated wave function using 
1400 lowest eigenstates determined by the numerical diagonalization of Eq.~(\ref{ro.10}) (black dashed line). 
Values on the axes are dimensionless, the time being scaled by the fundamental frequency $\omega _0$. 
Parameters of the Hamiltonian (\ref{ro.10}) are: $\eta =1.4$, $\alpha _{xx}=1.9\times 10^{-3}$, $\alpha _{xy}=1.0\times 10^{-3}$, 
and $\alpha _{yy}=1.1\times 10^{-3}$. The initial coherent state is characterized by 
$\langle x\rangle =4\sqrt{2}$, $\langle -i\partial /(\partial x)\rangle =0$, 
$\langle y\rangle =0$, $\langle -i\partial /(\partial y)\rangle =7\sqrt{\eta /2}$ at $t=0$.  
\label{fo.1}}
\end{caption}
\end{figure}

We test our method on the Hamiltonian of a two-dimensional (2D) harmonic oscillator with a quartic perturbation: 
\begin{eqnarray} 
\hat H&=&\hbar \omega_0\bigg[ \frac 12\left( -\frac {\partial ^2}{\partial x^2}-\frac {\partial ^2}{\partial y^2}+x^2+\eta ^2y^2\right) + 
\nonumber \\ & & 
\alpha _{xx}x^4+2\alpha _{xy}x^2y^2+\alpha _{yy}y^4\bigg]
.\label{ro.10} 
\end{eqnarray}
The co-ordinates $x$ and $y$ in Eq. (\ref{ro.10}) are dimensionless. 
Since this Hamiltonian describes only two modes, many its eigenstates $|\Phi _\lambda \rangle $ and respective eigenenergies 
$\epsilon _\lambda $ can be found numerically with a high precision up to pretty high values of $\epsilon _\lambda $.  
In Fig.~\ref{fo.1} we show the quantum mechanical 
mean value and variance of $x$ obtained by solving Eq.~(\ref{ro.6}) in compartison with the results directly following from the expansion  
$|\Psi (t) \rangle = \sum _\lambda e^{-i\epsilon _\lambda t/ \hbar } |\Phi _\lambda \rangle \langle \Phi _\lambda |\Psi (0)\rangle $.   
The same data for  $y$ as well as the covariance 
$\mathrm{Cov} \{ xy\} = \langle xy\rangle - \langle x\rangle \langle y\rangle $ are shown in Supplemental Material \cite{SM}. 
Our numerical method reproduces the behavior of the expectation values and second-order correlations quite well. 
As the initially coherent wave packet disperses 
in the anharmonic potential, its regular motion characterized by oscillating expectation values of the co-ordinates is damped and 
the co-ordinate variances reach their asymptotic values. 
We tested numerical energy conservation for our method and found  $\langle \Psi |\hat H|\Psi \rangle $ not exhibiting 
a systematic drift and deviating from its initial value by 1\% at maximum \cite{SM}. 

\begin{figure}[t] 
\begin{center} 
\includegraphics[width=\columnwidth]{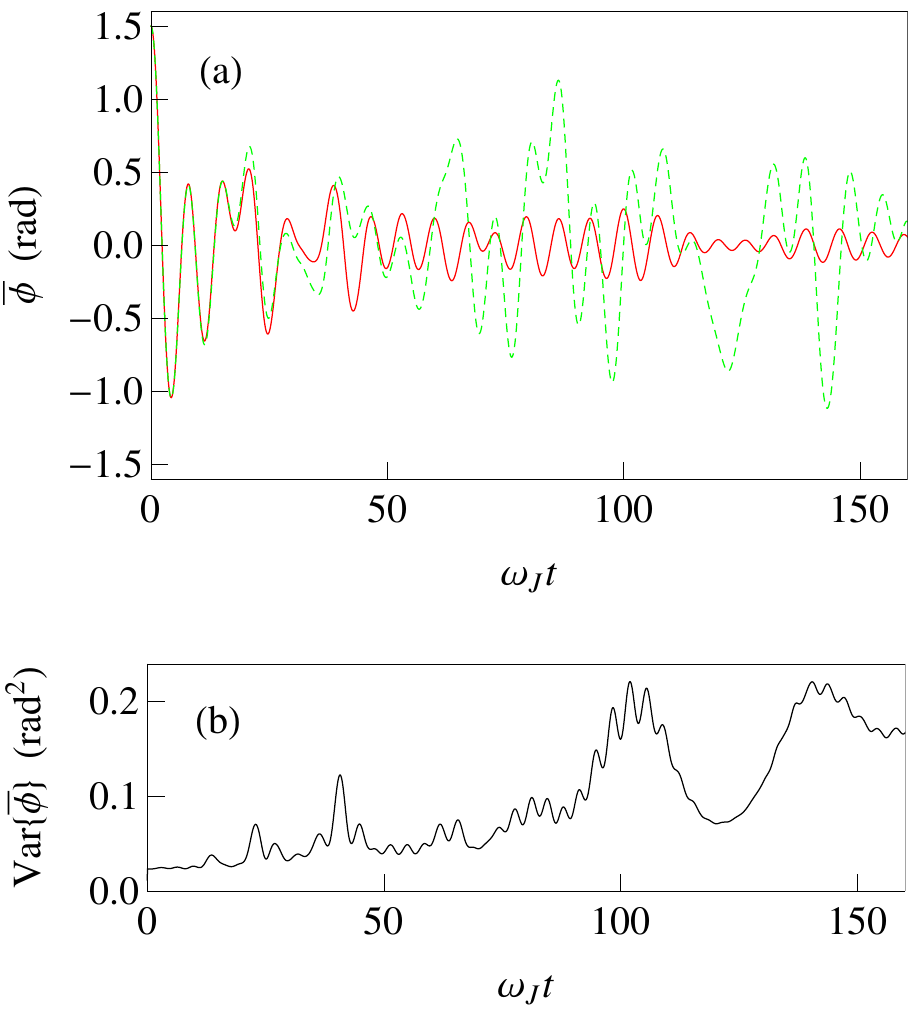}  
\end{center}

\begin{caption} 
{(Color online) (a) Mean global phase difference calculated using the quantum model Eq. (\ref{ro.6}) 
with ${\cal M}=21$ modes (red solid line) and the mean field approximation 
(green dashed line) and (b) its variance according to the quantum model as the function of the dimensionless time $\omega _\mathrm{J}t$ 
for $\hbar \omega _\Vert =3\, {U_\mathrm{c}}$ and $\hbar \omega _\mathrm{J} =30\, {U_\mathrm{c}}$. The initial coherent state 
corresponds to $\langle \hat \phi \rangle = 1.5 +0.5\sin (\pi s/2)$, $\langle \hat \rho \rangle = 0$. 
\label{fo.2}}
\end{caption}
\end{figure}

We choose the phase difference dynamics in an extended bosonic Josephson junction as the first application of our method to a 
system with a nontrivially large number of modes. We consider ultracold gas of bosonic atoms in a trap consisting of two 
tunnel-coupled atomic waveguides, to be referred as the left and right waveguides. 
The longitudinal trapping is harmonic with the fundamental frequency $\omega _\Vert $. In order to 
simplify the overview of the example, we make a few approximations. We assume that the number of atoms is small enough to neglect 
the dependence of the radial width of the atomic cloud on the local density \cite{Salasnich}, but large enough to provide 
an inverted parabolic Thomas--Fermi longitudinal profile of the 1D atomic density. Also we consider only antisymmetric modes 
(out-of-phase motion in the left and right waveguides) and neglect their coupling to the symmetric (in-phase) modes, thus 
reducing our problem to an ultracold-atom implementation of the sine-Gordon model, but, in contrast to Ref. \cite{polk-sg}, 
the mean 1D density profile is in our case non-uniform and proportional to $1-s^2$, where $s$ is the dimensionless (scaled to the 
Thomas--Fermi equilibrium radius $R_\mathrm{TF}$) longitudinal co-ordinate. The Hamiltonian reads  then 
\begin{eqnarray} 
\hat H &=&\int _{-1}^1ds\, \bigg[ \frac {U_\mathrm{c}}2\hat \rho ^2 + \frac {\hbar ^2 \omega _\mathrm{J}^2}{U_\mathrm{c}}
(1-s^2)\bigg( 1-\cos \hat \phi \bigg) + \nonumber \\ && 
\frac {\hbar ^2 \omega _\Vert ^2}{2U_\mathrm{c}} (1-s^2) \bigg( \frac {\partial \hat \phi }{\partial s}\bigg) ^{\! 2}\, \bigg ] . 
\label{ro.11} 
\end{eqnarray} 
Here $\hat \phi $ is the operator of the local phase difference between the left and right waveguides, $\hat \rho $ is the 
operator of the conjugate (density-difference) variable. The latter is made dimensionless by scaling to $R_\mathrm{TF}^{-1}$, since 
we use the dimensionless co-ordinate $s$, so that the 
commutation relation $[\hat \rho (s),\hat \phi (s^\prime )]=i\delta (s-s^\prime )$ holds. The charging energy $U_\mathrm{c}$ is positive, 
since we consider repulsive interactions between atoms. 
In repulsively interacting quantum gases at low energies the kinetic energy is dominated by phase fluctuations 
\cite{Dalfovo,ufn98}, therefore we omit 
a term proportional to $[\partial \hat \rho /(\partial s)]^2$ in Eq. (\ref{ro.11}) from the very beginning.  
The Josephson oscillation frequency corresponding to the peak mean density (at $s=0$) is denoted by $\omega _\mathrm{J}$. 

In Fig.~\ref{fo.2} the results of modeling of the mean and variance of the global phase difference 
$\bar \phi =\frac 12  \int _{-1}^1ds\, \hat \phi  $ are presented (we drop the operator hat above $\bar \phi $ to keep notation simple; 
this observable  can be measured  by standard experimental techniques \cite{Pigneur}). Regular Josephson oscillations decay and 
the quantum uncertainty of $\bar \phi $ becomes large compared to its zero-point level, but still within a range 
corresponding to high visibility of the integrated interference picture. Their damping of oscillations is not as fast and perfect as 
in the experiment \cite{Pigneur}, perhaps, because the model Hamiltonian (\ref{ro.11}) designed to demonstrate the proof of 
principle is too simplified.   

The numerical method to solve Eq. (\ref{ro.6}) 
is overviewed in Supplemental Materials \cite{SM}. Its most non-trivial part is related to the evaluation of 
exponential factors appearing in the r.h.s. of Eq. (\ref{ro.7}). On first glance one may get an 
impression that, e.g., estimation of a quartic interaction Hamiltonian for ${\cal M}$ modes requires independent calculation 
of ${\cal O}({\cal M}^4)$ different terms. However, this ``curse of dimensionality" \cite{curde} can be circumvented 
in a very efficient way. At  times  $g_{jj^\prime }t\lesssim 1$, which are sufficiently long 
to observe the collapse of coherence, we can make two simplifications. Firstly, 
we can neglect the non-commutativity of $\hat b_j$ and $\hat{\cal A}_{j^\prime }$  
and, hence, set $\nu ^{j_1 \dots j_l}_{j^\prime _1 \dots j^\prime _{l^\prime }}\approx \sum _{\kappa =1}^{l}\omega _l-
\sum _{\kappa ^\prime =1}^{l^\prime }\omega _{l^\prime }$ [see Eqs. (\ref{ro.3},\, \ref{ro.4},\, \ref{ro.8})]. 
Secondly, we can write 
$\exp [ -|\psi _j|^2 ( 1- e^{iG^{j_1 \dots j_l}_{j^\prime _1 \dots j^\prime _{l^\prime };j}t }) ]\approx 
\exp [ i|\psi _j|^2 G^{j_1 \dots j_l}_{j^\prime _1 \dots j^\prime _{l^\prime };j}t - 
\frac 12(|\psi _j|G^{j_1 \dots j_l}_{j^\prime _1 \dots j^\prime _{l^\prime };j}t)^2]$, employ the identity 
$\exp [-\frac 12(|\psi _j|G^{j_1 \dots j_l}_{j^\prime _1 \dots j^\prime _{l^\prime };j}t)^2]=
\int _{-\infty }^\infty d\sigma \, \exp (i\sigma |\psi _j|G^{j_1 \dots j_l}_{j^\prime _1 \dots j^\prime _{l^\prime };j}t
-\frac 12 \sigma ^2)/\sqrt{2\pi }$, and replace the integration by numerical averaging over a normally distributed pseudorandom 
parameter $\sigma $. These two approximations radically reduce rank of tensors used in numerical evaluation of the r.h.s. of 
Eq. (\ref{ro.6}).  
  
To summarize, we developed a novel approach to numerical simulation of the dynamics of finite-size bosonic systems 
beyond the mean field approximation. Our method is free from the curse of dimensionality and 
designed to evaluate time scales of the multimode quantum  
dynamics manifested through the collapse of coherence  
as well as expectation values and correlations of simple observables.  The description 
of scattering of quanta into initially empty modes remains beyond its scope. The main advantage of our method compared to 
the multiconfiguration approaches \cite{Cederbaum90,TC2008} is its remarkable simplicity and numerical efficiency in terms of 
the computational time and resources; it can be applied to obtain qualitative estimations on such long time scales 
that the number of configurations needed for the solution by standard methods becomes impractically large.  
Our method may be used not only in physics of 
trapped ultracold atomic gases, but also in other fields, such as molecular and chemical physics.  

The author thanks C. L\'ev\^eque, N. J. Mauser, J.-F. Mennemann, 
J. Schmiedmayer, and H.-P. Stimming for helpful discussion. This work is supported 
by  the Wiener Wissenschafts- und Technologie Fonds (WWTF) via Grant No. MA16-066 (SEQUEX).

\newpage

\onecolumngrid

\centerline{\textbf{Supplemental Material for}} 
\centerline{\textbf{Scrambled Mean Field Approach to the Quantum Dynamics of Degenerate Bose Gases}}

\vspace*{5mm} 

\centerline{Igor E. Mazets}  
\centerline{\textit{Vienna Center for Quantum Science and Technology, Atominstitut, TU~Wien,}} 
\centerline{\textit{Stadionallee~2,~1020~Vienna,~Austria; }} 
\centerline{\textit{Wolfgang Pauli Institute c/o Fakult\"{a}t f\"{u}r Mathematik,
Universit\"{a}t Wien,}} 
\centerline{\textit{ Oskar-Morgenstern-Platz 1, 1090 Vienna, Austria}}

\vspace*{5mm}

\twocolumngrid

\textbf{I. The 2D model} 
\vspace*{2mm}

For the 2D Hamiltonian Eq. (10) we introduce 
\begin{eqnarray} 
\hat b_x&=&\frac 1 {\sqrt 2}\left( x+\frac \partial {\partial x}\right) ,\nonumber \\
\hat b_x^\dag &=&\frac 1 {\sqrt 2}\left( x-\frac \partial {\partial x}\right) ,\nonumber \\
\hat b_y&=&\frac 1 {\sqrt 2}\left( \sqrt \eta \, y+\frac 1{\sqrt \eta }\frac \partial {\partial y}\right) ,\nonumber \\
\hat b_y^\dag &=&\frac 1 {\sqrt 2}\left( \sqrt \eta \, y-\frac 1{\sqrt \eta }\frac \partial {\partial y}\right) \nonumber 
\end{eqnarray} 
and evaluate the r.h.s. of Eq. (6) exactly. The obtained set of two equations for complex functions 
$\psi _x$ and $\psi _y$ is solved using a standard package of Wolfram \textsc{Mathematica 8}. The results for the 
expectation value and the variance of $y$ as well as for the covariance of $x$ and $y$ are plotted  below (red solid line). 
Black dashed line shows the results following from the numerical diagonalization of Eq. (10). 

\vspace*{2mm} 

\includegraphics[width=0.93\columnwidth]{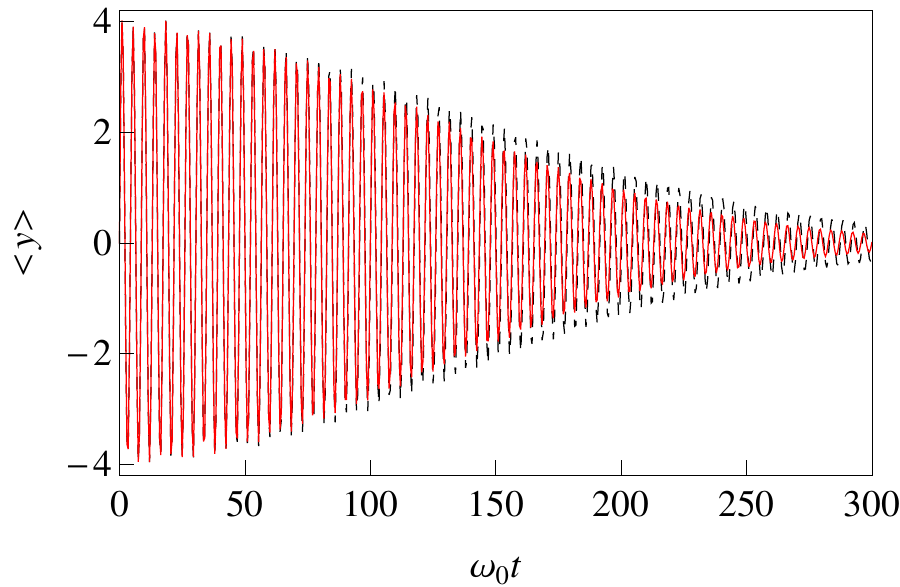} 

\vspace*{2mm} 

\includegraphics[width=0.93\columnwidth]{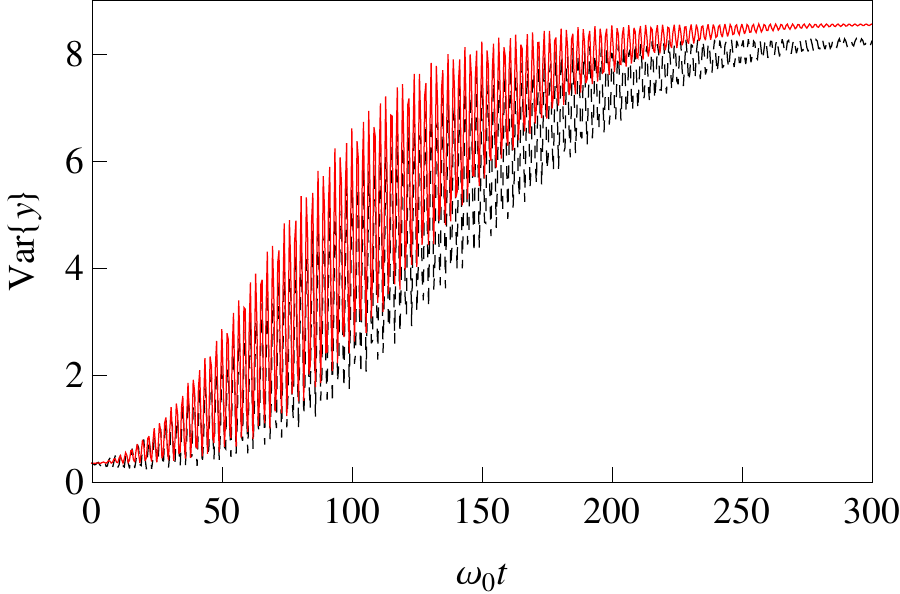}

\includegraphics[width=0.93\columnwidth]{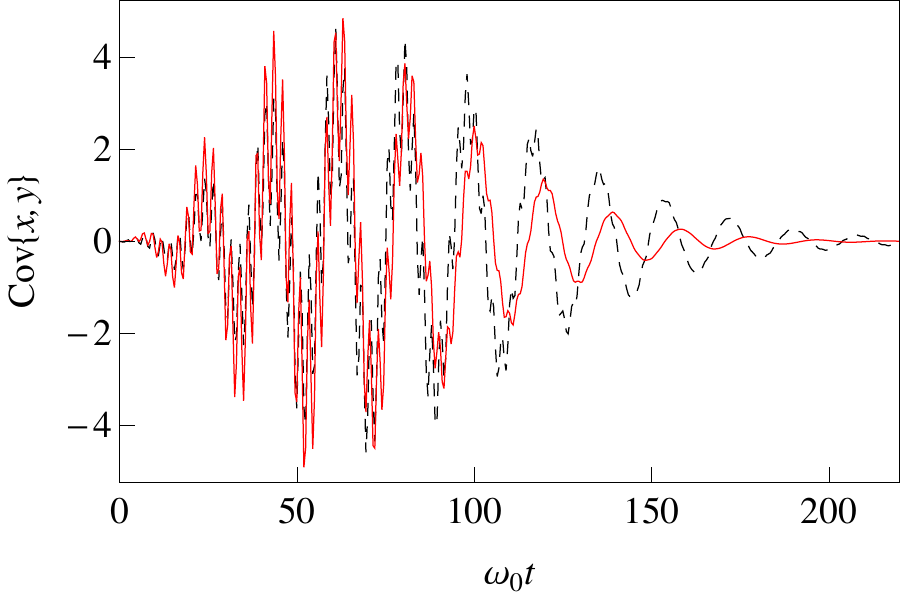} 

The energy conservation by our scrambled mean-field method is evident from the following ratio of the numerical mean energy $E_\mathrm{num}$
to the mean energy $E_\mathrm{ex}$ of the initial state calculated from the results of numerically exact diagonalization of Eq. (10): 

\vspace*{3mm} 

\includegraphics[width=0.95\columnwidth]{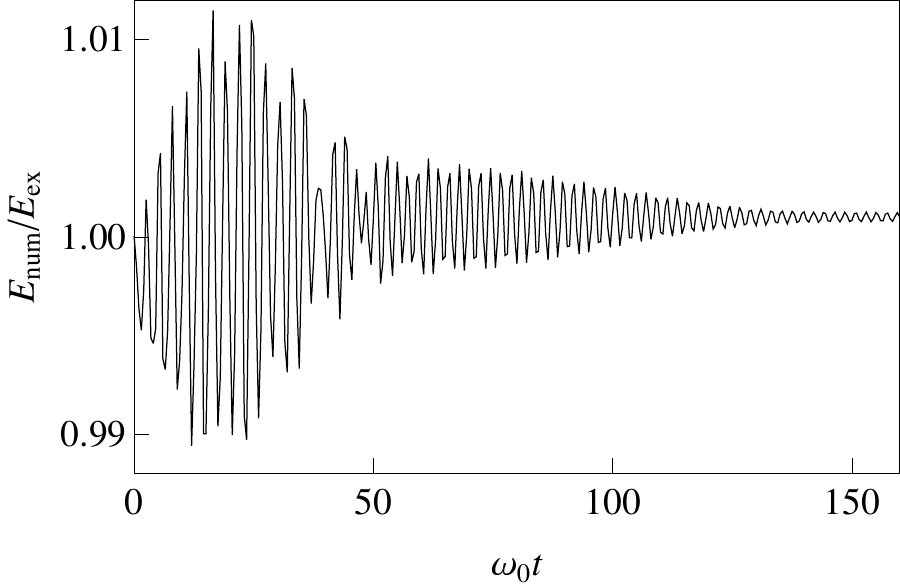} 

\vspace*{3mm} 

\textbf{II. Josephson junction} 

\vspace*{3mm}

The harmonic part of the Hamiltonian Eq. (11) is diagonalized by solving the eigenvalue problem 
$$
\omega _\mathrm{J}^2(1-s^2)f_j(s)-\omega _\Vert ^2\frac d{ds}\left[ (1-s^2)\frac {df_j(s)}{ds}\right] = \omega _j^2f_j(s), 
$$
with the boundary condition requiring $f_j\vert _{s\rightarrow \pm 1}$ to be finite. The functions $f_j(s)$ are real and 
orthonormalized, $\int _{-1}^1ds\, f_j(s)f_{j^\prime }(s)=\delta _{jj^\prime }$. Then the annihilation and creation operators 
are defined as 
\begin{eqnarray}  
\hat b_j      &=& \sqrt { \frac{\hbar \omega _j}{2U_\mathrm{c}} }\hat P_j +
i\sqrt { \frac{U_\mathrm{c}}{2\hbar \omega _j}{U_\mathrm{c}} }\hat \Phi _j , \nonumber    \\
\hat b_j^\dag &=& \sqrt { \frac{\hbar \omega _j}{2U_\mathrm{c}} }\hat P_j - 
i\sqrt { \frac{U_\mathrm{c}}{2\hbar \omega _j}{U_\mathrm{c}} }\hat \Phi _j , \nonumber   
\end{eqnarray}  
where 
$\hat P_j = \int _{-1}^1ds\, f_j \hat \rho $ and $\hat \Phi _j = \int _{-1}^1ds\, f_j \hat \phi $. 

To simulate numerically the evolution of the anharmonic system in the basis of ${\cal M}=21$ mode, we replaced 
the integral in Eq. (11) with a sum over ${\cal M}$ Gauss--Legendre quadrature points [1]. 
This quadrature formula approximates an integral of a function $z(s)$  as 
$\int _{-1}^1 ds\, z(s)\approx \sum _{k=1}^{\cal M}w_kz(s_k)$, where $s_k$'s are the roots of the Legendre polynomial   
$P_{{\cal M}}(s)$ and $w_k =2/[{\cal M}P_{{\cal M}-1}(s)\frac d{ds} P_{{\cal M}}(s)]\vert _{s=s_k}$ are the respective weights. 
%%The matrix $T_{kl} =\sqrt{l+\frac 12}P_l(s_k)\sqrt{w_k}$ is orthogonal, $\sum_{k=1}^{\cal M}T_{l^\prime k}T_{kl}=\delta _{l^\prime l}$. 
Then Eq. (11) in the ${\cal M}$-mode approximation reads as 
\begin{eqnarray} 
\hat H&=&\sum _{j=1}^{\cal M}\left( \frac{U_\mathrm{c}}2\hat P_j^2+\frac {\hbar ^2\omega _j^2}{2U_\mathrm{c}}\hat \Phi _j^2\right) + 
\nonumber \\ && 
\frac {\hbar ^2\omega _\mathrm{J}^2}{U_\mathrm{c}}\sum _{k=1}^{\cal M}w_k(1-s_k^2)
(1-\frac 12 \hat \phi _k^2-\cos \hat \phi _k), 
\nonumber 
\end{eqnarray} 
where the second line represents  the anharmonic part $\hat H_\mathrm{a}$ 
of the Hamiltonian and $\hat \phi _k= \sum_{j=1}^{\cal M}f_j(s_k)\hat \Phi _j$. 
The coupling constants in $\hat H_{\mathrm{qd}}$ are then given by 
$$
g_{jj^\prime }= -\frac {U_\mathrm{c}}{4\hbar (1+\delta _{jj^\prime })} \frac {\omega _\mathrm{J}^2}{\omega _j\omega _j^\prime } 
\sum _{k=1}^{\cal M}w_k(1-s_k^2)f_j^2(s_k)f_{j^\prime }^2(s_k).   
$$

As discussed in the main text, on the time scale $t\lesssim \hbar /U_\mathrm{c}$ we neglect the anharmonicity-induced corrections 
(of the order of $U_\mathrm{c}/\hbar $) to eigenfrequencies $\omega _j$. To the same approximation we neglect the difference 
between $\hat H_\mathrm{a}$ and its normally ordered form $:\hat H_\mathrm{a}:$, where the normal ordering is taken with respect to the 
creation and annihilation operators of collective excitations (but not of atoms). 

Eq. (6) is solved using the second-order predictor-corrector method. The main difficulty is to evaluate 
$\exp [ -|\psi _j|^2 ( 1- e^{iG^{j_1 \dots j_l}_{j^\prime _1 \dots j^\prime _{l^\prime };j}t }) ]\approx 
\exp [ i|\psi _j|^2 G^{j_1 \dots j_l}_{j^\prime _1 \dots j^\prime _{l^\prime };j}t - 
\frac 12(|\psi _j|G^{j_1 \dots j_l}_{j^\prime _1 \dots j^\prime _{l^\prime };j}t)^2]$ on each step. 
As explained in the main text, we 
use the identity $\exp (-\frac 12 \tau ^2) =\langle e^{i\sigma \tau }\rangle $, where $\langle \dots \rangle $ denotes averaging 
over Gaussian fluctuations of the random parameter $\sigma $ with zero mean and unity variance. This is implemented as follows. 
Before starting the numerical propagation in Eq. (6), we prepare the set of real vectors $\sigma ^{(j)} _\gamma $, $j, \gamma = 1, \, 2,
\, \dots \, , \, {\cal M}$ by generating ${\cal M}$ pseudorandom vectors, orthogonalizing them and imposing normalization 
$\sum _{\gamma =1}^{\cal M} \sigma ^{(j)} _\gamma \sigma ^{(j^\prime )} _\gamma =\delta _{jj^\prime }$. This normalization corresponds 
to $\langle \sigma ^{(j)\, 2} _\gamma \rangle =1$. The same set of vectors $\sigma ^{(j)} _\gamma $ is used throughout the 
numerical propagation of Eq. (6). 
Each time step of the predictor-corrector scheme is repeated $2{\cal M}$ times 
with a temporal replacement of $\psi _j$ by $\psi _j  \exp ( i\ell \sum _{j^\prime }g_{jj^\prime }|\psi _{j^\prime }| 
\sigma ^{(j^\prime )} _\gamma t)$ with $\gamma $ running from 1 to ${\cal M}$ and $\ell =\pm 1$ and the averaged result determines 
the values of $\psi _j$ at the next grid point of the time axis. Note that taking both signs of $\ell $ provides 
automatically the equality to zero of the imaginary part of our approximation of Gaussian factors. 

Our solution is stable against choosing different sets of $\sigma ^{(j)} _\gamma $. As we can see from the next plot, the discrepancy  
$\Delta \bar \phi $ between values of the global phase difference $\bar \phi $ obtained from two numerical solutions with 
different choices of $\sigma ^{(j)} _\gamma $ is of the order of $10^{-2}$~rad. 

\vspace*{3mm}

\includegraphics[width=0.95\columnwidth]{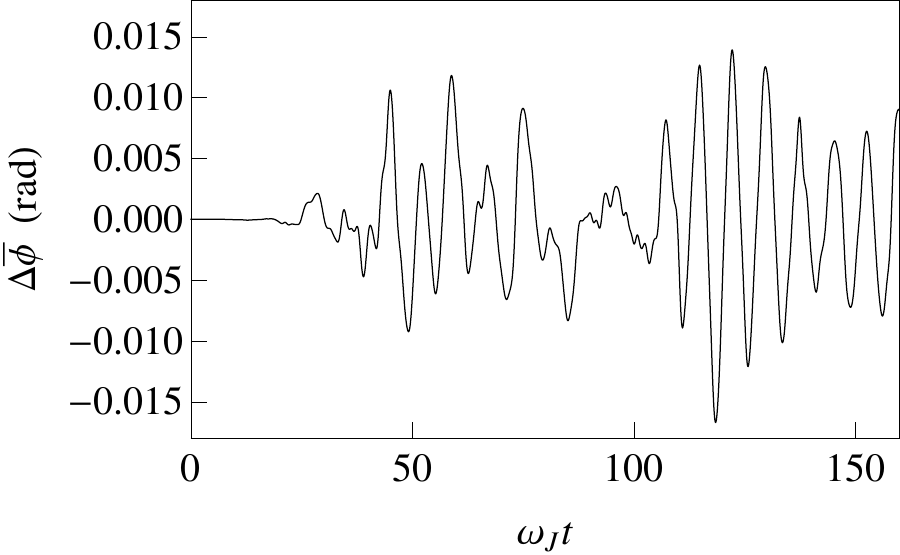} 

\vspace*{3mm} 

Our final test shows that the energy is conserved at the accuracy of about 2\%, as we can see from comparison of the averaged values 
of the Hamiltonian $\langle \hat H\rangle \equiv \langle \Psi |\hat H|\Psi \rangle $ at $t>0$ and $t=0$. 

\vspace*{1mm} 

\includegraphics[width=0.95\columnwidth]{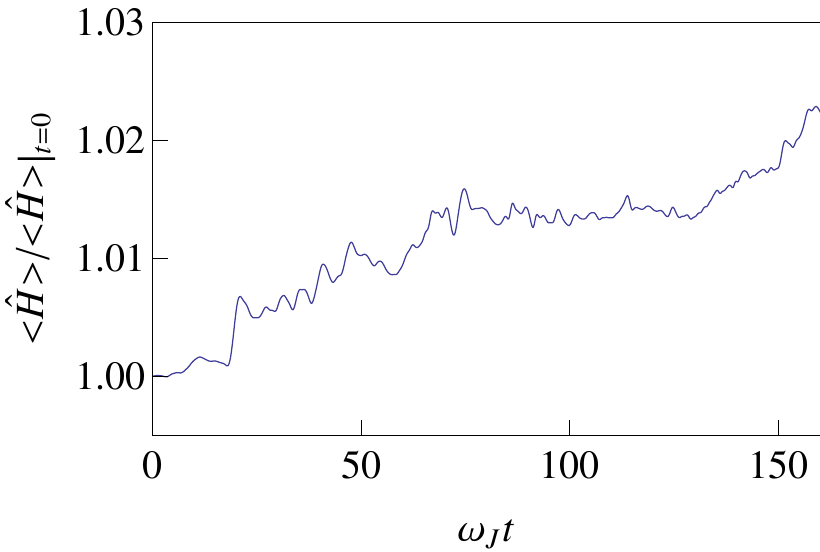} 

\vspace*{1mm} 

\centerline{***}
%%%%%\vspace*{1mm}

\noindent 
[1] {D. P. Laurie, J. Comput. Appl. Math. 
\textbf{127}, 201 (2001).}


\begin{thebibliography}{99} 
\bibitem{Dalfovo} F. Dalfovo, S. Giorgini, L. P. Pitaevskii, and S. Stringari, 
Theory of Bose-Einstein condensation in trapped gases. 
Rev. Mod. Phys. \textbf{71}, 463 (1999).

\bibitem{ufn98} L. P. Pitaevskii, Bose-Einstein condensation in magnetic traps. Introduction to the theory. 
Physics-Uspekhi \textbf{41}, 569   (1998).

\bibitem{Leggett-rev} A. J. Leggett, Bose-Einstein condensation in the alkali gases: Some fundamental concepts. 
Rev. Mod. Phys. \textbf{73}, 307 (2001). 

\bibitem{TWA1} M. J. Steel, M. K. Olsen, L. I. Plimak, P. D. Drummond, S. M. Tan, M. J. Collett, D. F. Walls, and R. Graham, 
\textit{Dynamical quantum noise in trapped Bose-Einstein condensates}.  
Phys. Rev. A \textbf{58}, 4824 (1998).

\bibitem{TWA2} A. Sinatra, C. Lobo, and Y. Castin, \textit{The truncated  Wigner method for Bose-condensed
gases: limits of validity and applications}. J. Phys. B \textbf{35}, 3599 (2002). 

\bibitem{Gasenzer1} T. Gasenzer, J. Berges, M. G. Schmidt, and M. Seco, 
Nonperturbative dynamical many-body theory of a Bose-Einstein condensate.  
Phys. Rev. A \textbf{72}, 063604 (2005). 

\bibitem{Werner} H. Aoki, N. Tsuji, M. Eckstein, M. Kollar, T. Oka, and P. Werner, 
Nonequilibrium dynamical mean-field theory and its applications.  
Rev. Mod. Phys. \textbf{86}, 779 (2014). 

\bibitem{Gasenzer2} I. Chantesana, A. P. Orioli, and T. Gasenzer, 
Kinetic theory of nonthermal fixed points in a Bose gas. 
Phys. Rev. A \textbf{99}, 043620 (2019). 

\bibitem{DiehlEPJD} M. Buchhold and S. Diehl, Kinetic theory for interacting Luttinger liquids. 
Eur. Phys. J. D  \textbf{69}, 224 (2015). 

\bibitem{Jin96} D. S. Jin, J. R. Ensher, M. R. Matthews, C. E. Wieman, and E. A. Cornell, 
Collective Excitations of a Bose-Einstein Condensate in a Dilute Gas.  
Phys. Rev. Lett. \textbf{77}, 420 (1996). 

\bibitem{Mewes96} M.-O. Mewes, M. R. Andrews, N. J. van Druten, D. M. Kurn, D. S. Durfee, C. G. Townsend, and W. Ketterle, 
Collective Excitations of a Bose-Einstein Condensate in a Magnetic Trap. 
Phys. Rev. Lett. \textbf{77}, 988 (1996). 

\bibitem{Alon08} O. E. Alon, A. I. Streltsov, and L. S. Cederbaum, 
Multiconfigurational time-dependent Hartree method for bosons: Many-body dynamics of bosonic systems.  
Phys. Rev. A \textbf{77}, 033613 (2008).

\bibitem{Hulet19} J. H. V. Nguyen, M. C. Tsatsos, D. Luo, A. U. J. Lode, G. D. Telles, V. S. Bagnato, and R. G. Hulet, 
Parametric Excitation of a Bose-Einstein Condensate: From Faraday Waves to Granulation. 
Phys. Rev. X 9, 011052 (2019). 

\bibitem{Takacs2} D. X. Horv\'ath and G. Tak\'acs, {Overlaps after quantum quenches in the sine-Gordon model}. 
Phys. Lett. B \textbf{771},  539  (2017). 

\bibitem{Takacs1} I. Kukuljan, S. Sotiriadis, and G. Takacs, 
{Correlation Functions of the Quantum Sine-Gordon Model in and out of Equilibrium}.  
Phys. Rev. Lett. \textbf{121}, 110402 (2018).

\bibitem{Bloch1} M. Greiner, O. Mandel, T. W. H\"ansch, and I. Bloch, 
Collapse and revival of the matter wave field of a Bose–Einstein condensate.  
Nature \textbf{419},  51 (2002).

\bibitem{Bloch2} S. Will, T. Best, U. Schneider, L. Hackerm\"uller, D.-S. L\"uhmann,   
and I. Bloch, Time-resolved observation of coherent multi-body interactions    
in quantum phase revivals. Nature \textbf{465}, 197 (2010).

\bibitem{Zhou}  T. Zhou, K. Yang, Z. Zhu, X. Yu, S. Yang, W. Xiong, X. Zhou, X. Chen, 
C. Li, J. Schmiedmayer, X. Yue, and Y. Zhai, 
Observation of atom-number fluctuations in optical lattices via quantum collapse and revival dynamics.  
Phys. Rev. A \textbf{99}, 013602 (2019). 

\bibitem{phase-diff} J. Javanainen and M. Wilkens, Phase and Phase Diffusion of a Split Bose-Einstein Condensate.  
Phys. Rev. Lett. \textbf{78}, 4675 (1997).

\bibitem{Imamoglu-theor} A. Imamo\={g}lu, M. Lewenstein, and L. You, 
Inhibition of Coherence in Trapped Bose-Einstein Condensates.   Phys. Rev. Lett. \textbf{78}, 2511 (1997).

\bibitem{Kuklov} A. B. Kuklov, N. Chencinski, A. M. Levine, W. M. Schreiber, and J. L. Birman,  
Quantum dephasing of normal modes of a Bose-Einstein condensate in a magnetic trap. 
Phys. Rev. A \textbf{55}, R3307(R) (1997). 

\bibitem{cum-collapse} F. W. Cummings, Stimulated emission of radiation in a single mode. 
Phys. Rev. \textbf{140},  A1051 (1965).

\bibitem{eber-collapse} J. H. Eberly, N. B. Narozhny, and J. J. Sanchez-Mondragon,  
Periodic spontaneous collapse and revival in a simple quantum model. 
Phys. Rev. Lett. \textbf{44}, 1323 (1980). 

%%%\bibitem{Salomaa} M. M\"ott\"onen, S. M. M. Virtanen, and M. M. Salomaa, 
%%%Collapse and revival of excitations in Bose-Einstein condensates. 
%%%Phys. Rev. A \textbf{71}, 023604 (2005).

%%\bibitem{GGE} T. Langen, S. Erne, R. Geiger, B. Rauer, T. Schweigler, M. Kuhnert, 
%%W. Rohringer, I. E. Mazets, T. Gasenzer, and J. Schmiedmayer, 
%%Experimental Observation of a Generalized Gibbs Ensemble, Science \textbf{348}, 207 (2015). 

\bibitem{Cederbaum87} J. Kucar, H.-D. Meyer, and L. S. Cederbaum, Time-dependent rotated Hartree approach. 
Chem. Phys. Lett. \textbf{140}, 525 (1987). 

%%%%\bibitem{Salomaa} M. M\"ott\"onen, S. M. M. Virtanen, and M. M. Salomaa, 
%%%%Collapse and revival of excitations in Bose-Einstein condensates. 
%%%%Phys. Rev. A \textbf{71}, 023604 (2005).

\bibitem{Glauber} R. J. Glauber,  Coherent and Incoherent States of the Radiation Field.  
Phys. Rev. \textbf{131}, 2766 (1963).  

\bibitem{SM} See Supplemental Material. 

\bibitem{Salasnich} L. Salasnich, A. Parola, and L. Reatto, 
Effective wave equations for the dynamics of cigar-shaped and disk-shaped Bose condensates.  
Phys. Rev. A \textbf{65}, 043614 (2002).

\bibitem{polk-sg} V. Gritsev, A. Polkovnikov,  and E.  Demler,  
Linear response theory for a pair of coupled one-dimensional condensates of interacting atoms. 
Phys. Rev. B, \textbf{75},  174511  (2007).´

\bibitem{Pigneur} M. Pigneur, T. Berrada, M. Bonneau, T. Schumm, E. Demler, and J. Schmiedmayer, 
Relaxation to a Phase-Locked Equilibrium State in a One-Dimensional Bosonic Josephson Junction.
Phys. Rev. Lett. \textbf{120}, 173601 (2018).

\bibitem{curde} J. F. Traub and  A. G. Werschulz, \textit{Complexity and information} 
(Cambridge University Press, Cambridge,  1998). 

\bibitem{Cederbaum90} H.-D. Meyer. U.Manthe, and L. S. Cederbaum, 
{The multi-configurational time-dependent Hartree approach}. Chem. Phys. Lett. \textbf{165}, 73 (1990).

\bibitem{TC2008} J. M. Bowman, T. Carrington, and H.-D. Meyer, Variational quantum approaches for computing vibrational 
energies of polyatomic molecules. Molecular Phys. \textbf{106}, 2145 (2008). 
\end{thebibliography}
\end{document}